\documentclass[twocolumn,showpacs,preprintnumbers,amsmath,amssymb]{revtex4}
\usepackage{graphicx}%
\usepackage{enumerate}

\def\be{\begin{equation}}
\def\ee{\end{equation}}
\def\bee{\begin{eqnarray}}
\def\eee{\end{eqnarray}}

\begin{document}
\title{Toroidal momentum transport in a tokamak caused  by symmetry breaking parallel derivatives}

\author{T.~Sung$^1$, R.~Buchholz$^1$, F.J.~Casson$^2$, E.~Fable$^2$, S.R.~Grosshauser$^1$, W.A.~Hornsby$^1$, P.~Migliano$^1$, A.G.~Peeters$^1$}

\vskip 0.2 truecm 

\address{$^1$ University of Bayreuth, Physics department, Universit\" atsstrasse 30 Bayreuth, Germany}
\address{$^2$ Max Planck Institut fuer Plasmaphysik, EURATOM association, Boltzmannstrasse 2, 85748 Garching, Germany}

\pacs{52.25.Fi, 52.25.Xz, 52.30.Gz, 52.35.Qz, 52.55.Fa}

\begin{abstract}
A new mechanism for toroidal momentum transport in a tokamak is investigated using the gyro-kinetic model. 
First, an analytic model is developed through the use of the ballooning transform.    
The terms that generate the momentum transport are then connected with the poloidal derivative of the ballooning envelope, which are one 
order smaller in the normalised Larmor radius, compared with the derivative of the eikonal.
The mechanism, therefore, does not introduce an inhomogeneity in the radial direction, in contrast with the effect of profile shearing. 
Numerical simulations of the linear ion temperature gradient mode with adiabatic electrons, retaining the finite $\rho_*$ effects in the ExB velocity, 
the drift, and the gyro-average, are presented. 
The momentum flux is found to be linear in the normalised Larmor radius ($\rho_*$) but is, nevertheless, generating a 
sizeable counter-current rotation.  
The total momentum flux scales linear with the aspect ratio of the considered magnetic surface, and increases with increasing magnetic shear, 
safety factor, and density and temperature gradients. 
\end{abstract}

\maketitle 

\section{INTRODUCTION} 

For a tokamak, plasma rotation can be beneficial for confinement and stability. 
In particular, if the rotation is associated with a sufficiently large ExB shear, it stabilises turbulence \cite{BIG90,WAL95,STA13}, and 
leads to an enhanced energy and particle confinement.  
For this reason, there is a strong interest in the experiments on, and the theoretical description of, plasma rotation. 
An observation of particular relevance to a fusion reactor is the occurrence of spontaneous rotation \cite{eri97,ric98a, ric99, hut00, ric00, hoa00, deg04, sak06, sca06, bor06, ric07}, i.e.~a plasma rotation that develops without an externally applied torque. 
The theoretical description of this spontaneous rotation has attracted much interest, and significant progress has been made in recent years. 
For an overview we refer the reader to Ref.~\cite{PEE11}. 

All contributions to toroidal momentum transport are associated with a parity symmetry breaking mechanism in the gyro-kinetic equation \cite{PEE05,PEE11,PAR11}. 
It was shown in \cite{PEE11} that to lowest order in the normalised Larmor radius $\rho_* = \rho_i / R_0 \ll 1$, where $\rho_i$ is the ion Larmor 
radius and $R_0$ is the major radius of the magnetic axis, a finite momentum flux is generated only in the presence of a rotation
gradient \cite{MAT88} or 
for a rotating plasma through the Coriolis pinch \cite{pee07,hah07} or for a non up-down symmetric magnetic equilibrium \cite{cam09b,cam09}.  
In next order many different mechanisms appear. 
The most studied mechanisms to day are: the effect of neo-classical equilibrium flows \cite{PAR10}, 
the ExB shear \cite{dom93,gar02,gur07,wal07,cas09} and some form of radial profile shearing \cite{CAM11,GUR10,WAL11}. 
The latter two mechanisms introduce an inhomogeneity in the radial direction. 

In this paper a new mechanism of toroidal momentum transport is studied. 
The mechanism is connected with higher order, in the normalised Larmor radius $\rho_*$, parallel derivatives of the perturbed distribution as 
well as the perturbed fields.  
For micro-instabilities, the perturbed quantities have scale lengths perpendicular to the magnetic field that are of the order of the Larmor radius. 
The fast transport along the magnetic field, however, strongly damps short scale perturbations, and the scale length along the magnetic field is of 
the order of the system size ($R_0$). 
This ordering allows some parallel derivatives that appear in the model equations to be neglected since they are one order smaller in the normalised 
Larmor radius ($\rho_*$) compared with the perpendicular gradients. 
The neglected terms, however, do break the parity symmetry and might, despite their smallness, be of relevance in the description 
of momentum transport. 
In this paper the effect of these parallel derivatives on momentum transport is studied. 
The paper concentrates on quasi-linear theory, leaving the non-linear state for further study. 

Before turning to the complete model description and its numerical solution, the physics mechanism is discussed below using an analytic model. 
We stress that this model does not include all contributions to the momentum flux obtained in the model that is developed in the next section. 
The goal of the analytic model is to clarify the physics and to provide a first estimation of the expected momentum transport. 
The physics mechanism considered in this paper can be most easily clarified using the ballooning transform with $\hat s-\alpha$ geometry \cite{CON78}. 
All perturbed quantities are then assumed to have the form 
\be 
G(r,\theta,\varphi) = \hat G(\theta) \exp [ {\rm i} n ( \varphi - q(r) \theta ) - {\rm i} \omega t] + {\rm c.c.}, 
\label{balloon} 
\ee
where $\varphi$ ($\theta$) is the toroidal (poloidal) angle, $r$ is the radius of the flux surface, $n \gg 1$ is the toroidal mode number of the instability, $q$ is the safety factor, 
$\omega = \omega_R + {\rm i} \gamma$ is the complex frequency, and c.c.\space denotes the complex conjugate. 
The rapid variation ($n \gg 1$) of the perturbation perpendicular to the magnetic field is represented by the eikonal, with the argument ($\varphi - q \theta$) chosen such that it is constant 
along the magnetic field. 
The envelope $\hat G$ is then assumed a slowly varying function of the poloidal angle ($\theta$). 
The ExB velocity (${\bf v}_E = {\bf b} \times \nabla \phi / B$) using the expression above is
\be 
\hat {\bf v}_{E} = {{\rm i} n \over B} {\bf b} \times \nabla ( \varphi - q \theta) \hat \phi(\theta) + {\bf b} \times \nabla \theta {1 \over B} {\partial \hat \phi \over \partial 
\theta}, \label{ExB}
\ee
where $\phi$ is the electro-static potential, ${\bf b}$ is the unit vector along the magnetic field ${\bf B}$, $B$ is the magnetic field strength, and the FLR effects connected with the gyro-average 
of the potential have been neglected for the sake of clarity. 
The first term in the expression above is due to the gradient of the eikonal, whereas the second term is due to the derivative of the envelope. 
For micro-instabilities, $n \propto 1/ \rho_* \gg 1$. 
The second term is one order smaller in the normalised Larmor radius compared with the first and is, therefore, neglected in the lowest order local limit.
In this paper the term 'lowest order local limit' will be used to indicate the gyro-kinetic model that retains terms only to lowest relevant order in the normalised Larmor radius.
In this limit the radial variation of plasma and geometry parameters is neglected, i.e. the model is local to a specific magnetic surface. 
The new physics, finite $\rho_*$, terms introduced in this paper do not introduce a radial inhomogeneity and, therefore, the model remains 'local', but not 'lowest order'.
The discussed physics effects are therefore distinct from any form of profile shearing, which relies on the introduction of a radial inhomogeneity.  
In the lowest order local limit the solution of the gyro-kinetic equation yields a potential perturbation that is symmetric in $\theta$ \cite{PEE05,PEE11,PAR11}. 
It can be directly verified that the first term in Eq.~(\ref{ExB}) then yields a contribution that is symmetric in $\theta$, whereas
the second term is anti-symmetric. Based on this  model we are searching for higher order $\rho_*$ terms that breaks the parity
symmetry of gyro-kinetic equation
The higher order $\rho_*$ contribution to the ExB velocity breaks the parity symmetry of the gyro-kinetic equation and leads to a finite, non-diffusive, flux of toroidal momentum.

A simple estimate of the consequences of the higher order $\rho_*$ contribution in the ExB velocity on the momentum flux can be obtained by considering 
the gyro-kinetic equation for singly charged ions retaining only the effect of the fluctuating ExB velocity in the background
Maxwell distribution, and the 
acceleration along the magnetic field due to the electro-static potential  
\be 
{\partial f \over \partial t} = - {\bf v}_E \cdot \nabla F_M +  {e \over m} {\bf b} \cdot \nabla \phi 
{\partial F_M \over \partial v_\parallel}, 
\ee
where $v_\parallel$ is the parallel velocity, $e$ the elementary charge, $m$ is the ion mass, $f$ the perturbed ion distribution
function, and $F_M$ the Maxwellian of the ions
\be 
F_M = {n_0 \over \pi^{3/2} v_{th}^3} \exp \left [ - {(v_\parallel - R B_t \omega_\varphi/ B)^2 + 2 \mu B / m \over v_{th}^2 } \right ] .
\ee
In the equation above $n_0$ is the particle density, $v_{th} = \sqrt{2T/m}$ the thermal velocity, $\omega_\varphi(r)$ is the radial profile of the
angular toroidal rotation frequency, $B_t$ the toroidal component of the magnetic field, $R$ the major radius, $\mu = m v_\perp^2 / 2 B$ the magnetic moment, $T$ the temperature, and $v_\perp$ the velocity perpendicular to the magnetic field. 
Building the parallel velocity moment of the (reduced) gyro-kinetic equation, assuming a Maxwell closure (for the 
moments see Ref.~\cite{pee09c}), yields an equation for the perturbed parallel velocity ($\hat w$)    
\be 
- {\rm i} \omega \hat w  = -{{\rm i} k_\theta \over B} \left [ \hat \phi + {{\rm i} \over nq} 
{\partial \hat \phi \over \partial \theta} \right ] R_0 {\partial \omega_\varphi \over \partial r} 
- {e \over m} {1 \over qR_0} {\partial \hat \phi \over \partial \theta} ,
\label{weq}
\ee 
where $k_\theta = n q / r$, and ${\bf b} \cdot \nabla \hat \phi = (1 / qR_0) \partial \hat \phi / \partial \theta$
have been used.  
Furthermore, the gradient of the Maxwellian has been evaluated at the surface for which $\omega_\varphi = 0$ (i.e.~there is no Coriolis pinch contribution), and finite inverse aspect ratio effects have been neglected ($R B_t / B \approx R_0$, ${\bf b} \approx {\bf e}_\varphi$).

The radial flux of toroidal momentum averaged over the flux surface is 
\be 
\Gamma_\varphi^r = {1 \over 4 \pi^2} \oint {\rm d} \varphi {\oint} {\rm d} \theta \, m n_0 v_E^r w ,
\ee
where $v_E^r$ is the radial component of the ExB velocity, and finite inverse aspect ratio effects in the flux surface average have been 
neglected. 
Substituting the eikonal form of Eq.~(\ref{balloon}) and integrating over the toroidal angle yields 
\be \label{eq:Gamma_varphi}
\Gamma_\varphi^r = {1 \over 2 \pi} \oint {\rm d} \theta \, {\rm Re} \left ( 2 m n_0 \hat v_E^r \hat w^\dagger
\right ) ,
\ee
where Re denotes the real part and the dagger the complex conjugate. 
From the equation of the perturbed parallel velocity, using the complex frequency $\omega = \omega_R + {\rm i} 
\gamma$ one can derive 
\bee
\hat w^\dagger &=& {\omega_R + {\rm i} \gamma \over \vert \omega \vert^2 } {k_\theta \over B} 
\left [ \hat \phi^\dagger - {{\rm i} \over nq} {\partial \hat \phi^\dagger \over \partial \theta} \right ] 
R_0 {\partial \omega_\varphi \over \partial r} \cr 
\noalign{\vskip  0.2 truecm} 
& &+ {\rm i} {\omega_R + {\rm i} \gamma \over \vert \omega\vert^2}
{ e \over m} {1 \over q R_0} {\partial \hat \phi^\dagger \over \partial \theta} .
\eee
Then substituting the expression for the ExB velocity, and retaining the lowest order relevant $\rho_*$ terms only, 
one obtains 
\bee 
{\rm Re} [ \hat v_E^r \hat w^\dagger ] &=& {\gamma \over \vert \omega \vert^2} \left [ 
-{k_\theta^2 \over B^2} \vert \phi \vert^2 R_0 {\partial \omega_\varphi \over \partial r} + {k_\theta  e \over q R_0 B m} 
{1 \over n q}  \left \vert {\partial \hat \phi \over \partial \theta }\right \vert^2 \right ] \cr 
\noalign{\vskip  0.2 truecm} 
& & - {\rm Re} \left [ { e k_\theta \over q R_0 B m} {\omega_R + {\rm i} \gamma \over \vert \omega \vert^2} 
\hat \phi {\partial \hat \phi^\dagger \over \partial \theta} \right ] .
\eee
To proceed the potential amplitude is split in a real and imaginary part 
\be 
\hat \phi = \hat \phi_R + {\rm i}\hat \phi_I. 
\ee
Using the periodicity constraint in the poloidal angle, one obtains 
\be 
\oint {\rm d} \theta \hat \phi {\partial \hat \phi^\dagger \over \partial \theta} = 
2{\rm i} \oint {\rm d} \theta \, \hat \phi_I {\partial \hat \phi_R \over \partial \theta} .
\ee

At this point a particular choice for the potential $\hat \phi$ must be made. 
If no symmetry breaking mechanisms are present the potential is symmetric in $\theta$. 
However, the higher order $\rho_*$ contribution to the ExB velocity breaks the symmetry and, 
since the ExB velocity directly contributes to the drive of the instability, we assume a potential 
that has a form similar to the radial component of the ExB velocity  
\be 
\hat \phi = \hat \phi_0 + {{\rm i} \over nq} {\partial \hat \phi_0 \over \partial \theta} ,
\label{phieq}
\ee 
where $\hat \phi_0$ is symmetric in $\theta$.
We note that a finite momentum flux is found from the equations given above even if we choose $\hat \phi 
= \hat \phi_0$. 
This is because finite $\rho_*$ terms have been kept in the equation for the perturbed velocity $\hat w$ (Eq.~(\ref{weq})).
The second term on the right hand side of Eq.~(\ref{phieq}) that represents the broken symmetry for the potential and is 
$\rho_*$ smaller than the first term, further enhances the momentum flux.   
The asymmetry in the potential, proposed in the equation above, is directly observed in the numerical simulations 
that will be discussed in Section III. 

Substituting the expression for the potential given by Eq.~(\ref{phieq}), and using the symmetry of $\hat \phi_0$, again 
considering the lowest order relevant $\rho* \propto 1/n$ terms, one obtains
\be 
\oint {\rm d} \theta \hat \phi {\partial \hat \phi^\dagger \over \partial \theta} = 
{2 {\rm i} \over n q} \oint {\rm d} \theta \left \vert {\partial \hat \phi_0 \over \partial \theta } \right \vert^2 .
\ee 
Using this expression in the equation for the momentum flux we finally obtain 
\bee 
\Gamma_\varphi^r &=& {m n_0 k_\theta^2 T^2 \over \pi e^2 B^2} {\gamma \over \vert \omega \vert^2} \oint {\rm d} \theta \, 
\biggl [ -\left \vert {e \hat \phi_0 \over T} \right \vert^2 R_0 {\partial \omega_\varphi \over \partial r} +\cr 
\noalign{\vskip 0.2 truecm}  
& &+ {3 v_{th} \rho_* \over q R_0 (k_\theta \rho)^2 \epsilon} \left \vert {\partial \over \partial \theta}\left ({e \hat \phi_0 \over T} 
\right ) \right \vert^2 \biggr ] ,
\eee
where the Larmor radius is $\rho = m v_{th} / eB$. 
If no external torque is applied on the plasma the equation above predicts a stationary rotation gradient $u^\prime_{EQ} = - (R_0^2/ v_{th}) (\partial 
\omega_\varphi / \partial r)$ 
\be
u^\prime_{EQ} = - {3 \over q (k_\theta \rho)^2 \epsilon } 
{\oint \vert \partial (e \hat \phi_0 / T)/\partial \theta \vert^2 \, {\rm d} \theta \over 
\oint \vert e \hat \phi_0 / T \vert^2 \, {\rm d} \theta } \rho_* .
\ee
The sign of the gradient means that the plasma centre would rotate in the counter-current direction, if the rotation 
frequency at the edge is zero. 
The ratio of the integrals can be estimated by introducing a scale length $L_\theta$ for the potential perturbation 
\be\label{uprim_analytic} 
u^\prime_{EQ} = - {3  \over q (k_\theta \rho)^2 \epsilon L_\theta^2}  \rho_* . 
\ee 
The effect is obviously a finite $\rho_*$ effect, but for standard parameters $q = 1.4$, $k_\theta \rho = 0.3$, $\epsilon = 0.16$,
and $L_\theta = 1$ one obtains 
\be 
u^\prime_{EQ} = - 149 \rho_* .
\ee
The estimate above indicates that for the DIII-D cyclone base case \cite{DIM00} $\vert u^\prime_{EQ} \vert = 0.38$
even without any external momentum input. 
This value is to be compared with the range $\vert u^\prime_{EQ} \vert = [0-1]$ obtained in H-mode \cite{PEE05} with 
neutral beam heating and, consequently, with an applied external torque. 
An alternative method to estimate its relevance is to compare this contribution with the more familiar ExB shear. 
Adopting the estimate of the ExB shear rate $\gamma_E =  \rho_* (R/L_T)^2$ (for details see Ref.~\cite{PEE11}) one obtains for the cyclone case  
$\vert u^\prime_{EQ}\vert = 47\rho_*$. 
The mechanism discussed here can dominate over the ExB shearing but, of course, the relative strength depends on the actual plasma parameters considered. 
Even though the momentum flux introduced in this section is due to a finite $\rho_*$ correction, it is far from negligible. 
When developing the $\rho_*$ expansion, all quantities like $\epsilon$ and $k_\theta \rho$ are taken to be of order unity, but 
the products of such quantities can still result in a factor comparable to $1/\rho_*$.  

The mechanism discussed above is only one of several obtained when considering derivatives towards the parallel coordinate. 
Following the discussion above we can identify several finite $\rho_*$ effects that 
might lead to substantial momentum flux contributions: 
\begin{enumerate}[A] 
\item Higher order $\rho_*$ corrections to the ExB velocity. (The effects of this mechanism have been estimated above.)
\item Higher order $\rho_*$ corrections in the treatment of the drift due to the magnetic field inhomogeneity (${\bf v}_D$)
\item Higher order $\rho_*$ corrections to the particle trapping. 
\item Higher order $\rho_*$ corrections in the gyro-average as well as polarisation. (Note that these two are interlinked.) 
\item Higher order $\rho_*$ correction to the calculation of the fluxes. (This is linked with the first point) 
\end{enumerate} 
It will be shown below that each of these effects breaks the parity symmetry of the gyro-kinetic equation, and leads to momentum transport. 

\section{THE MODEL} 
\newcommand{\g}[2]{g^{#1#2}}
\newcommand{\D}[1]{\mathcal D^{#1}}
\newcommand{\E}[2]{\mathcal E^{#1#2}}
\newcommand{\I}[1]{\mathcal I^{#1}}
\newcommand{\G}{ {B^s\over B^2}{\partial B \over \partial s}}
\newcommand{\F}{ {B^s\over B} }
\renewcommand{\L}{\mathcal L}
\renewcommand{\H}[1]{\mathcal H^#1}

Starting point of the model used in this paper is the gyro-kinetic equations of motion 
\bee
{{\rm d} {\bf X} \over {\rm d} t} &=& v_\parallel {\bf b}+{\bf v}_D+{\bf v}_E \cr 
\noalign{\vskip 0.2 truecm} 
{{\rm d} v_\parallel  \over {\rm d} t} &=&  - { 1 \over m v_\parallel} {{\rm d}{\bf X} \over {\rm d} t} \cdot 
\left(Z e\nabla\langle\phi\rangle+\mu\nabla B\right),
\label{equations_of_motion} 
\eee
where ${\bf v}_D$ is the drift velocity due to the inhomogeneous magnetic field, and ${\bf v}_E$ is the ExB drift,
\bee
{\bf v}_D&=& { 1 \over Ze }\left( {mv_\parallel^2 \over B} + \mu \right) {{\bf B}\times\nabla B\over B^2}+ \frac{mv_\parallel^2}{2Ze}
\beta^\prime\frac{ {\bf B}\times \nabla \psi}{B^2} \cr 
\noalign{\vskip 0.2 truecm} 
{\bf v}_E&=&{{\bf b}\times \nabla \langle\phi\rangle\over B}.
\eee
In this paper we do not consider the drift due to the Coriolis and Centrifugal forces \cite{pee09c,CAS10}. 
In Eq.~(\ref{equations_of_motion}) the angle brackets around the potential ($\phi$) indicate the gyro-average 
\be
\label{gyro_av}
\langle\phi\rangle=\frac{1}{2\pi} \oint d\alpha\, \phi({\bf R}+{\boldsymbol \rho}) ,
\ee
where $\alpha$ is the gyro-angle, $\rho$ is the Larmor radius, and   
\be
{\boldsymbol\rho} = \rho ({\bf e}_1 \cos \alpha + {\bf e}_2\sin\alpha) 
\label{rho_vec}
\ee
is the vector pointing from the gyro centre to the particle position. 
The vectors ${\bf e}_1$ and ${\bf e}_2$ are orthogonal unity vectors perpendicular to the magnetic field (${\bf B}$).
The equations above are not correct to all orders in the normalised Larmor radius \cite{BRI07,IDO12}.  
The model, therefore, is unable to retrieve all finite $\rho_*$ corrections. 
It is, however, able to describe the finite $\rho_*$ effects mentioned in the introduction as it will be shown below. 

Field aligned Hamada coordinates $(\psi, \zeta, s)$ are used. For these coordinates the contra-variant components of the 
magnetic field are flux functions and $B^\psi = B^\zeta = 0$. 
The radial coordinate $\psi$ is a flux label, and the coordinates are chosen such that $\zeta$ remains an ignorable 
coordinate, i.e.~all scalars that satisfy toroidal symmetry are not a function of $\zeta$ (For instance $B = B(\psi,s)$).
The coordinate $s$ acts as a parallel coordinate ${\bf B} \cdot \nabla = B^s {\partial / \partial s}$. 
The perturbed distribution function as well as the perturbed fields are assumed to have a scale length of the order 
$\rho$ in the $\psi$ and $\zeta$ direction, but a scale length $R_0$ along the magnetic field. 
This leads to the ordering 
\be \label{eq:ordering}
{\cal O}\biggl({\partial f \over \partial s}\biggl) = \rho_* {\cal O}\biggl ({\partial f \over \partial \psi},\rho_*{\partial f \over \partial 
\zeta}   \biggr ) .
\ee
It must be noted here that, although $s$ acts as a parallel coordinate, $\nabla s$ does not point along the magnetic 
field. 
In fact, in the simplified $\hat s-\alpha$ geometry $s = \theta / 2 \pi$ and $\nabla s$ points in the poloidal direction. 
The higher order $\rho_*$ derivatives towards $s$ that are neglected in the lowest order local limit, therefore, are  
similar to the derivatives of the envelope in the ballooning transform. 
Below we will go through the various effects mentioned in the introduction and discuss what changes to the local 
model are necessary to retain them. 
The momentum fluxes are calculated using the gyro-kinetic code GKW, and for details on the local model the reader 
is referred to Ref.~\cite{PEE09b}.

\subsection{Higher order $\rho_*$ corrections to the ExB velocity}

In the linear theory the convection due to the ExB velocity is kept only for the background Maxwell distribution ($F_M$)  
\be
{\bf v}_E \cdot \nabla F_M  = {{\bf b} \over B} \cdot ( \nabla x^\alpha \times \nabla \psi  ) {\partial \langle \phi \rangle
\over\partial x^\alpha} {\partial F_M \over \partial \psi} ,
\ee
where $x^1 = \psi$, $x^2 = \zeta$, $x^3 = s$, and the Einstein summation convention has been applied. Due to the 
ordering, in the lowest order local limit, the derivative of $\langle \phi \rangle$ towards $s$ is neglected as it results in 
a term that is smaller by one order in the normalised Larmor radius. The additional term, not considered in the lowest order
local limit, that we have to add to the model in order to describe the finite $\rho_*$ effects therefore is 
\be
{\bf v}_E \cdot \nabla F_M  \,\,{\buildrel +\over =} \,\,{{\bf b} \over B} \cdot ( \nabla s \times \nabla \psi  ) {\partial \langle \phi \rangle
\over\partial s} {\partial F_M \over \partial \psi} .
\ee
Here, and below, we have used the symbol ${\buildrel +\over =}$ to indicate the finite $\rho_*$ terms, that have to be considered 
additionally to those of the lowest order local limit. 
Because the cross product of the gradients of the coordinates often appears in the equations, it is useful to define the tensor 
\be 
{\cal E}^{\alpha \beta} = {{\bf b} \over 2B} (\nabla x^\alpha \times \nabla x^\beta) ,
\ee
where the factor 2 has been introduced to make the definition equivalent with Ref.~\cite{PEE09b}. Then 
\be 
{\bf v}_E \cdot \nabla F_M \, {\buildrel + \over =} \, 2 {\cal E}^{s \psi} {\partial \langle \phi \rangle \over \partial s} 
{\partial F_M \over \partial \psi}. 
\ee

\subsection{Higher order $\rho_*$ correction in the drift due to the magnetic field inhomogeneity}

A similar argument applies to the convection caused by the drift velocity 
\be 
{\bf v}_D \cdot \nabla = v_D^\alpha {\partial \over \partial x^\alpha} .
\ee
Because the derivative of any perturbed quantity towards $s$ is one order smaller in the normalised Larmor radius, 
such derivatives are neglected in the lowest order local limit. Therefore the additional finite $\rho_*$ term that has to be 
considered is 
\bee
{\bf v}_D \cdot \nabla &{\buildrel +\over =}& v_D^s {\partial \over \partial s} = \cr
\noalign{\vskip 0.2 truecm} 
& &2 \left [  \left ( {mv_\parallel^2 \over ZeB} + {\mu \over Ze} \right ) {\partial B \over \partial \psi} + 
{mv_\parallel^2 \over 2 Z e} \beta^\prime \right ] {\cal E}^{\psi s}  {\partial \over \partial s}\hspace{2ex}. 
\eee

\subsection{Higher order $\rho_*$ corrections to the particle trapping}

In developing the lowest order local limit only the lowest order relevant $\rho_*$ terms are retained. 
The particle trapping is then evaluated considering the parallel convection only. 
For a finite beta plasma there is, however, a finite $\rho_*$ correction to the trapping connected with the drift motion 
\bee 
{\partial f \over \partial t} &{\buildrel + \over =}& {\mu v_\parallel \over 2 Z e B^2} \beta^\prime ({\bf B} \times \nabla \psi) 
\cdot \nabla B  {\partial f \over \partial v_\parallel} \cr 
\noalign{\vskip 0.2 truecm} 
&= & {\mu v_\parallel \over Z e} \beta^\prime {\cal E}^{\psi s} {\partial B \over \partial s} {\partial f \over 
\partial v_\parallel} .
\eee
This correction is not only small in $\rho_*$, it is also proportional to $\beta^\prime$ and, therefore, negligible for low $\beta$
experiments. 

\subsection{Higher order $\rho_*$ corrections in the gyro-average as well as polarisation} 

The gyro-average is defined through Eq.~(\ref{gyro_av}), with the vector ${\boldsymbol\rho}$ given by Eq.~(\ref{rho_vec}). 
The unit vectors in the latter equation can be taken to be 
\begin{align}\label{eq:e1}
{\bf e}_1&= {\nabla \psi\over \sqrt{\left|\nabla\psi\right|}} &   {\bf e}_2&= {{ \bf b} \times\nabla \psi\over
\sqrt{\left|\nabla\psi\right|}} .
\end{align}
The variation of the potential in the $s-$direction is small 
\be
{\boldsymbol \rho} \cdot\nabla s {\partial \phi \over \partial s}\sim {\mathcal O}\left(\rho_* {\boldsymbol \rho} \cdot\nabla
\psi { \partial \phi \over\partial \psi}\right), 
\ee
and the gyro-average in the lowest order local limit is performed considering the dependence of $\phi$ on $\psi$ and $\zeta$ only. 
In principle the full gyro-average can be performed by integrating over the ring retaining also the dependence of $\phi$ on $s$. 
However, since the variation with $s$ is small it is easier to use a Taylor expansion of $\phi$ in $s$, retaining only terms 
up to the first order 
\begin{multline}
\langle\phi\rangle(\psi,\zeta,s)=\frac{1}{2\pi} \oint d\alpha\,\biggl[ \phi(\psi+{\boldsymbol \rho}\cdot\nabla\psi,\zeta+{\boldsymbol
\rho}\cdot\nabla\zeta,s)\\
\left.+{\partial\phi(\psi+{\boldsymbol \rho}\cdot\nabla\psi,\zeta+{\boldsymbol\rho}\cdot\nabla\zeta,s)\over \partial s}{ \boldsymbol \rho}\cdot\nabla s\right]. 
\end{multline}
Because of the ordering the second term in the square brackets of the equation above is one order smaller in $\rho_*$. 

All simulations that retain the finite $\rho_*$ terms in the gyro-average, are performed using finite difference in the radial and $s-$direction 
while a spectral representation is used for the $\zeta$-direction. 
The integral of the gyro-average is performed using 32 points on the gyro-ring. 
The values of the function in between the radial grid points is obtained by linear interpolation. 
The derivative towards $s$ in the equation above is calculated using central differencing on the s-grid. 

Modifications in the gyro-average directly affect the gyro-kinetic Poison equation: 
\bee
& & - {Z^2 e^2 \over T} \int {\rm d}^3 {\bf v} \, ( \langle \langle \phi \rangle \rangle - \phi ) F_{M} 
\cr 
\noalign{\vskip 0.2 truecm}
& & = Z e \int {\rm d}^3 {\bf v} \, \langle f \rangle  - n_0 {e \phi \over T_e} 
\eee
where $Z$ is the ion charge number, $T_e$ the electron temperature, and the last term on the right hand side represents the 
adiabatic electron response. 
In this equation the gyro-average of both the perturbed ion distribution function $f$ as well as the gyro-average of the
gyro-averaged potential appears. 
Consistency demands that when the gyro-average of the potential in the evolution equation is modified, the same modification 
is applied to the Poisson equation.

\subsection{Higher order $\rho_*$ terms in calculating the fluxes. }

The quasi-linear toroidal momentum flux is evaluated as 
\be 
\Gamma_\varphi^\psi  = \left \{ \int {\rm d}^3 {\bf v} \, ({\bf v}_E \cdot \nabla \psi)  { m v_\parallel R B_t \over B} f  \right \}
,
\ee
where the brackets $\{\}$ denote the flux surface average. Consistent with the finite $\rho_*$ correction of ${\bf v}_E \cdot \nabla F_M$, 
a correction to toroidal momentum flux appears 
\be 
\Gamma_\varphi^\psi \,\, {\buildrel + \over =} \,\, \left \{ 2\int {\rm d}^3 {\bf v} \, {\cal E}^{s\psi} {\partial \langle \phi 
\rangle \over \partial s} {m v_\parallel R B_t \over B} f \right \}, 
\ee 
with similar expression for the fluxes of particles and energy.

\subsection{Model set of equations}

For the sake of completeness, and to document exactly which equations are being solved, in this section the full set of equations is 
given. 
The notation, and normalisation are given in Ref.~\cite{PEE09b}, and the reader is referred to this paper for further details. 
The evolution equation for the perturbed distribution $f$ consists of several contributions 
\be\label{eq:evolution}
{\partial f \over \partial t} = {\rm I} + {\rm II} +{\rm III} + {\rm IV} + {\rm V} + {\rm VI}
\ee
The various terms in this equation are 
\begin{align}
{\rm I} =& - v_\parallel {\bf b} \cdot \nabla f \rightarrow
- v_R v_{\parallel} {\F} {\partial \hat f \over \partial s} \label{model1} \\
{\rm II} =& - {\bf v}_D \cdot \nabla f \rightarrow 
-{{\rm i} k_p \over Z}  T_R E_D {\mathcal D}^p \hat f  + \notag\\
&\phantom{- {\bf v}_D \cdot \nabla f \rightarrow }
-\underbrace{\rho_*{1 \over Z}  T_R E_D {\mathcal D}^s {\partial \hat f\over\partial s} }_{B_1}\\
{\rm III} =& + {1 \over m v_\parallel} (v_\parallel {\bf b} + {\bf v}_D) \cdot\mu \nabla B  {\partial f \over \partial v_\parallel} \rightarrow\\
&v_R  B{B^s\over B^2} \mu {\partial f \over \partial v_\parallel} +
\underbrace{\rho_* \frac{T}{Z} v_\parallel {\mathcal E} ^ {\psi s} \frac{\partial B}{\partial s} \frac{\beta^\prime}{B^2}\mu {\partial f \over \partial
v_\parallel}}_{C}\\
{\rm IV} =& - {\bf v}_{E} \cdot \nabla F_M \rightarrow\notag\\
&\hspace{1em}{\rm i}k_p  {\langle\hat\phi\rangle} {\mathcal E}^{ p \psi} \left [ {1 \over L_n} +  E_T {1 \over L_T} + {2 v_{\parallel} \over v_R} 
{R B_t \over B} u^\prime \right ] F_{M} \notag\\
& +\underbrace{\rho_* {\partial{\langle\hat\phi\rangle}\over\partial s}{\cal E}^{s\psi} \biggl [ {1 \over L_n} + E_T {1 \over L_T} + 
{2 v_{\parallel} \over v_R}{ R B_t \over B} u^\prime \biggr ] F_{M}}_{A},\\
{\rm V} =& - {Z e \over T} v_\parallel {\bf b} \cdot \nabla \langle\hat \phi \rangle F_M \rightarrow - {Z \over T_R} v_R v_{\parallel} 
\F {\partial {\langle\hat \phi \rangle} \over \partial s} F_{M} ,\\
{\rm VI} =& - {Z e \over T}{\bf v}_D \cdot \nabla \langle\hat \phi \rangle F_M  \rightarrow 
- {\rm i}  E_D {\cal D}^p  k_p \, {\langle \hat\phi \rangle} \, F_{M} +\notag\\
&\phantom{- {Z e \over T}{\bf v}_D \cdot \nabla \langle \phi \rangle F_M  \rightarrow }
 - \underbrace{\rho_* E_D {\cal D}^s \, {\partial{\langle\hat \phi \rangle}\over\partial s} \, F_{M}}_{B_2} \label{model6},
\end{align}
where the tensor ${\cal D}$ is related to ${\cal E}$ through 
\be 
{\cal D}^\alpha = - 2 {\cal E}^{\alpha \beta} {1 \over B} {\partial B \over \partial x^\beta} 
\ee
and $E_T$ is
\be
E_T=v_\parallel^2+2\mu B-{3\over2}
\ee
Because we do not consider the non-linearity, or the neo-classical transport, the numbering of the terms is different from Ref.~\cite{PEE09b}. 
Also compared with Ref.~\cite{PEE09b} we consider only the electro-static case. 
The Latin index $p$ indicates a summation over $p = 1$, 2 only, i.e.~excluding the parallel direction.  
The higher order $\rho_*$ corrections introduced in this paper are marked with under-braces. Their labels correspond to the
enumeration of the symmetry breaking mechanism in the introduction, if the drift velocity in the perturbed ($B_1$) and background
($B_2$) distribution is considered as the mechanism $B$. 

Since $\rho_*$ is a small parameter, one can expect that some linearisation of the solution around $\rho_* = 0$ is possible. 
This implies that the momentum flux generated by the newly introduced terms in the Eqs.~(\ref{model1}-\ref{model6}) is linear in $\rho_*$. 
It also suggests that the total momentum flux generated by the finite $\rho_*$ corrections is the sum of the momentum fluxes 
generated by each of the $\rho_*$ terms individually, since their interaction would scale as $\rho_*^2$. 
Both these hypotheses will be tested in the next section.

\subsection{Symmetry breaking} 

The parity symmetry discussed in \cite{PEE05,PAR10,PEE11}. For the linear case considered here the transformation
\be 
v_\parallel \to -v_\parallel \quad
s \to  -s \quad 
\psi \to  -\psi \quad
\ee
leaves the gyro-kinetic equation invariant. Note that the transformation above implies for the wave vectors: $k_\psi \to -k_\psi$, 
$k_\zeta \to + k_\zeta$. 
Assuming an up-down symmetric equilibrium, the tensors ${\cal E}$ and ${\cal D}$ can 
be shown to have the following properties 
\begin{align}
\E \psi \zeta (s) & = +\E\psi \zeta(-s)  & \E \psi s(s) &  = +\E \psi s(-s) \\
\E \zeta s(s)& = -\E\zeta s(-s) & {\cal D}^\psi(s) & = - {\cal D}^\psi(-s) \\
{\cal D}^\zeta(s) & =+{\cal D}^\zeta(-s) & {\cal D}^s(s) & = +{\cal D}^s (-s)
\end{align}
It can then be directly verified that none of the terms retained in the lowest order local limit changes sign under the transformation 
given above. 
With only these terms kept in the model equations, the solution for the potential is symmetric in the low field side 
position, while the parallel velocity perturbation is anti-symmetric. 
The resulting momentum flux is then zero. 
It can however, also be verified that all finite $\rho_*$ terms do change sign. 
All these terms will break the symmetry and can generate a finite momentum flux.

\section{Results}

\begin{figure}[htb]
\centering
\includegraphics[width=\linewidth]{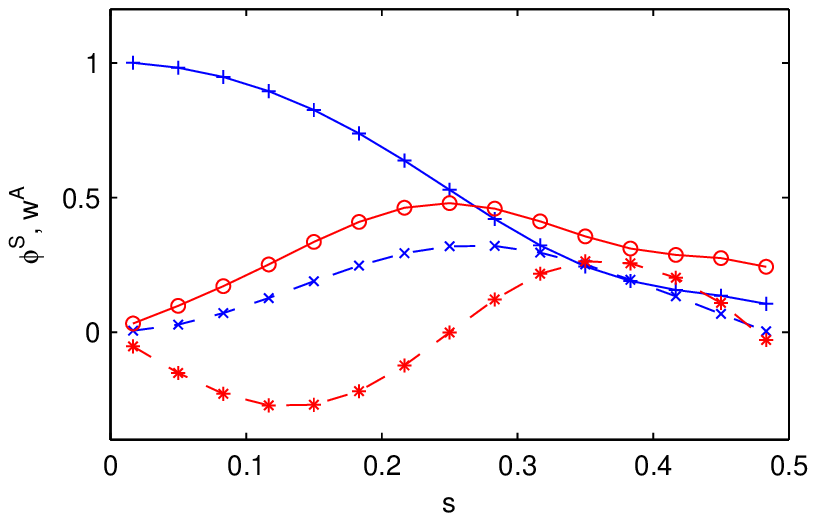}
\includegraphics[width=\linewidth]{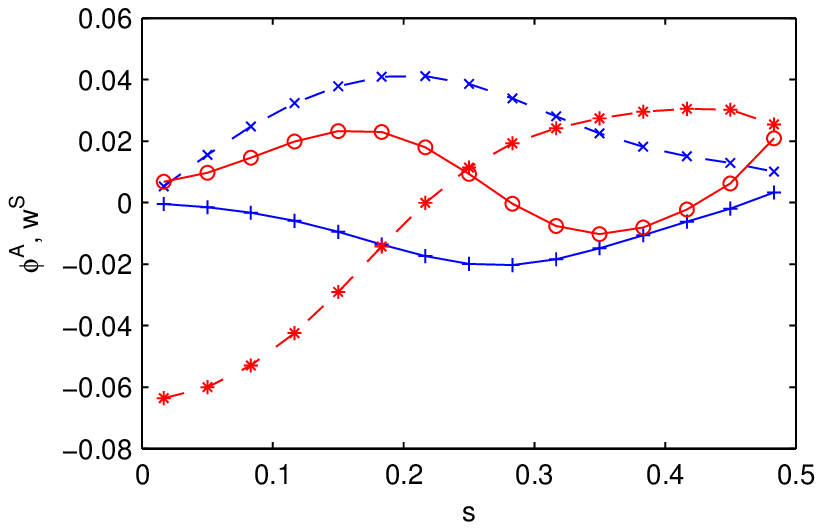}
\caption{(Color online) ${\rm Re}[\phi]$ [solid (blue) line with symbol '+'], ${\rm Im}[\phi]$ [dashed (blue) line with symbol 'x'], 
${\rm Re}[w]$ [solid (red) line with symbol 'o'], and ${\rm Im}[w]$ [dashed (red) line with symbol '*']. The top figure gives 
$\phi^S$ and $w^A$, while the bottom figure gives $\phi^A$, $w^S$. All functions shown in the lower figure are zero in the absence of finite 
$\rho_*$ terms.}
\label{potential}
\end{figure}

All simulation results in this paper are obtained with the gyro-kinetic code GKW \cite{PEE09b}, using a magnetic 
equilibrium of concentric circular surfaces where the full dependence on inverse aspect ratio $\epsilon$ is kept \cite{LAP09},
rather than the $\hat s-\alpha$ geometry where only the lowest order in $\epsilon$ is kept.
For this equilibrium, $\beta^\prime = 0$, and the correction to the trapping (mechanism C in the previous section) will not be 
considered. 
All of the simulations have been obtained using finite difference in the radial and $s-$direction, and a spectral representation 
for the $\zeta$-direction. 
Periodic boundary conditions are used in the radial direction, consistent with the homogeneous nature of the model. 
The chosen parameters are those of the Waltz standard case \cite{WAL94}: ion temperature gradient length $R/L_T = 9$, density
gradient length $R/L_N = 3$, magnetic shear $\hat s = 1$, safety factor $q = 2$, inverse aspect ration $\epsilon = 0.16$, ion 
to electron temperature ratio $T_i / T_e = 1$. 
The adiabatic electron approximation is employed, and we consider the linear stability and quasi-linear fluxes only.  
For the chosen parameters, the most unstable mode is the ion temperature gradient (ITG) mode. 
Unless otherwise specified, the normalised Larmor radius $\rho_*$ is $2.5\cdot 10^{-3}$, and poloidal wave vector $k_\theta \rho$ is  0.43.
Note that the definition of $\rho_*$ is different by a factor $\sqrt{2}$ compared with most literature due to a factor 2 the in thermal velocity 
$v_{th} = \sqrt{2 T / m}$, i.e. $\rho= \sqrt{2} \rho_s$ with $\rho_s = \sqrt{m_i T_e}/eB$. 
Furthermore, in $\rho_*$ the Larmor radius is normalised with the major radius ($R_0$) rather than the minor radius of the last 
closed flux surface ($a$).
For $a / R = 1/3$, $a/\rho_s = 188$ which is close to the value of the cyclone base case \cite{DIM00}.   
Simulations use 30 points in the $s-$direction, 16 $\mu$ and 16 $v_\parallel$, grid points.
Finally, 41 radial points with a box size of 20 $\rho$ are used in the radial direction. 
The representation on a radial grid, in contrast to the ballooning representation discussed in the introduction, corresponds to 
a range of radial wave vectors rather than one mode with a radial wave vector that is zero at the low field side of the torus.

Without any of the $\rho_*$ terms in Eqs.~(\ref{model1}-\ref{model6}) the potential ($\phi$), density and temperature perturbations are 
symmetric in the low field side position $s = 0$, whereas the parallel velocity perturbation ($w$) is anti-symmetric. 
When the finite $\rho_*$ terms are introduced the symmetry is broken for all quantities. 
The effect is clearly visible on, for instance, $\phi(s)$ but can nevertheless be better displayed by constructing the symmetric ($G^S$)
and anti-symmetric ($G^A$) component of every perturbed quantity 
\be 
G^S = {1 \over 2} \left [ G(s) + G(-s) \right ]  \qquad 
G^A = {1 \over 2} \left [ G(s) - G(-s) \right ] .
\label{symasym}
\ee
Fig.~\ref{potential} shows the real and imaginary part of the potential ($\phi$) and parallel velocity perturbation ($w$) decomposed in 
their symmetric and antisymmetric parts. 
Shown is the eigenfunction averaged over the radial domain as a function of $s \approx \theta / 2 \pi$, calculated considering all finite 
$\rho_*$ terms discussed in the previous section. 
The functions in the lower figure, $\phi^A$ and $w^S$, are zero in the absence of finite $\rho_*$ terms. 
It can be seen that the finite $\rho_*$ terms lead to a modest asymmetry in the potential and parallel velocity perturbations.  

\begin{figure}[htb]
\centering
\includegraphics[width=\linewidth]{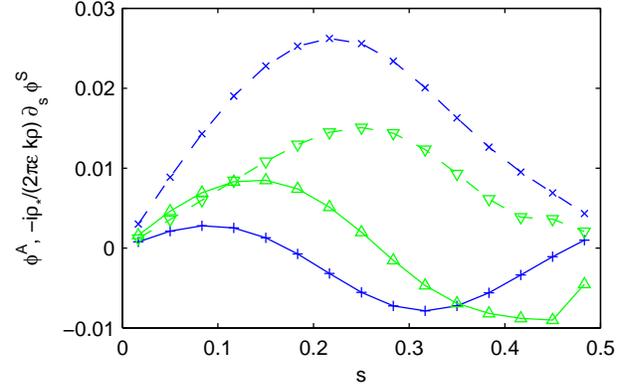}
\caption{(Color online) ${\rm Re}[\phi^A]$ calculated directly from the eigenfunction [solid (blue) line with symbol '+'], 
${\rm Im}[\phi^A]$ calculated directly from the eigenfunction [dashed (blue) line with symbol 'x'], ${\rm Re}[\phi^A]$ 
calculated through numerical differentiation of $\phi^S$ [solid (green) line with symbol '$\triangle$'], and ${\rm Im}[\phi^A]$ 
calculated through numerical differentiation of $\phi^S$ [dashed (green) line with symbol '$\triangledown$']}
\label{potential2}
\end{figure}

The eigenmode, together with the decomposition of Eqs.~(\ref{symasym}), can be used to verify the assumption represented 
by Eq.~(\ref{phieq}) of the analytic model. 
This assumption gives the following relation between the symmetric and anti-symmetric part of the potential  
\be 
\phi^A=  {i \over nq} {\partial \phi^S \over \partial \theta}. 
\ee
Fig.~\ref{potential2} shows the anti-symmetric part of the potential calculated directly from the eigenfunction, as well as 
through numerical differentiation of the symmetric part. 
Here, only the finite $\rho_*$ correction in the ExB velocity is retained in the numerical simulation (i.e. only contribution A of 
Eqs.~(\ref{model1}-\ref{model6}) is kept), in agreement with the assumptions made in the analytic model.  
Comparing the curves it can be seen that the agreement is reasonable, though not perfect.
This verifies that the analytic model gives a reasonable estimate of the effect. 
Numerical simulation is nevertheless necessary if accuracy is required. 

In this paper we consider two contributions to the momentum flux only: the diagonal (diffusive) contribution and the flux due to 
the finite $\rho_*$ terms. 
As discussed in the previous section, due to the smallness of $\rho_*$, one can expect the various mechanisms
to generate a momentum flux that is linear in $\rho_*$. 
The equation for the momentum flux may then be written in the form  
\be\label{eq:Gamma_linear}
\Gamma_\varphi^\psi  =  \chi_\varphi u^\prime + C \rho_* 
\ee
where $u^\prime = - R_0^2 \nabla \omega_\varphi / v_{th}$. 
The first term in the equation above is the momentum diffusivity which has been studied in several papers \cite{MAT88,PEE05,STR08}. 
In the absence of an external torque, the momentum flux due to the finite $\rho_*$ parallel derivatives will modify the rotation profile, 
until this flux is balanced by the diffusive momentum flux, and $\Gamma_\varphi^\psi = 0$. 
The equilibrium toroidal rotation gradient then is 
\be 
u^\prime_{EQ} = - {C \over \chi_\varphi} \rho_* 
\ee
The equation above provides a useful alternative to express the magnitude of the finite $\rho_*$ momentum flux, as it 
more clearly indicates its impact on the rotation profile. 

\begin{figure}[htb]
\centering
\includegraphics[width=\linewidth]{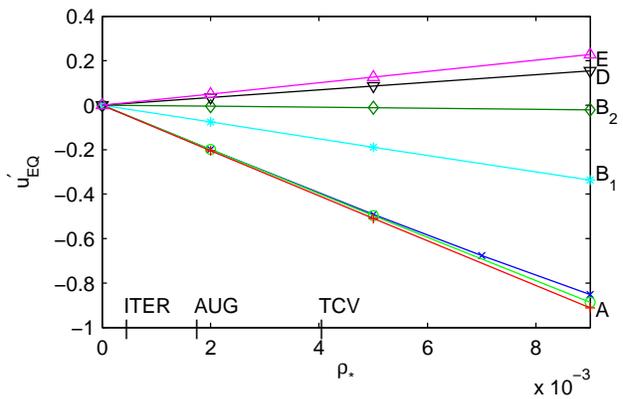}
\caption{(Color online) The momentum flux expressed in the equilibrium $u^\prime_{EQ}$ as a function of $\rho_*$.
The letters on the right relate to the symmetry breaking mechanism identified in Eq.~(\ref{model1}-\ref{model6}) 
A: red line with symbol '+', B$_1$: light blue line with symbol '*', B$_2$: dark green with symbol '$\diamond$', D: black line with
symbol '$\triangledown$', E: magenta line with symbol '$\triangle$'. Additionally the sum of all
individual contributions is given by the (green) curve with the symbol 'o', and 
the momentum flux calculated including all finite $\rho_*$ terms is given by the (blue) curve with symbol 'x'. The $\rho_*$ axis
contains typical values of three machines: ITER, AUG and TCV .
} 

\label{sum}
\end{figure}

Not only the flux due to the finite $\rho_*$ terms can be expected to be linear in $\rho_*$, the smallness of $\rho_*$ also 
suggests that the various effects are additive. 
Both these properties are shown in Fig.~\ref{sum}, which gives $u^\prime_{EQ}$ as a function of $\rho_*$ for each of the 
mechanisms discussed in the previous section, as well as the sum of all these separate contributions and the momentum 
flux calculated including all mechanisms. 
It can be seen that the flux due to all mechanisms is nearly perfect linear in $\rho_*$. 
For $\rho_*=0.005 (0.009)$ the momentum flux calculated including all mechanisms is 1\%(5\%) smaller than the sum of the 
contributions calculated for each of the mechanisms separately. 
The various symmetry breaking mechanisms therefore do not have a strong interaction.
The finite $\rho_*$ terms only have a small influence on the growth rate of the instability and the quasi-linear ion heat flux. 
For $\rho_*=0.005 (0.009)$ the growth rate increases by $0.4\%(1.4\%)$ and the ion heat flux decreases by 0.9\%(3\%).

It can be seen from the figure that the largest contribution to the momentum transport is the fluctuating ExB velocity in the background 
gradient, i.e.~the mechanism that was investigated in the introduction through an analytic model. 
Furthermore, $C \approx 100$ for the results shown in Fig.~\ref{potential}, in reasonable agreement (but larger than) the 
analytic estimate $C = 3 / (q (k_\theta \rho)^2 \epsilon L_\theta^2) = 51$ (using $L_\theta \approx 1$). 
The effect of the drift in the perturbed distribution on the momentum flux, as well as the corrections to the gyro-average and the calculation 
of the flux are smaller, though non-negligible, compared with the effect of the ExB velocity. 
For the parameters used in the simulations the sum of all effects is close to the contribution of the ExB velocity, due to a cancellation 
of the other contributions. 
This is, however, a coincidence for this particular set of parameters. 

\begin{figure*}[tb]
\centering
\includegraphics[width=0.3\linewidth]{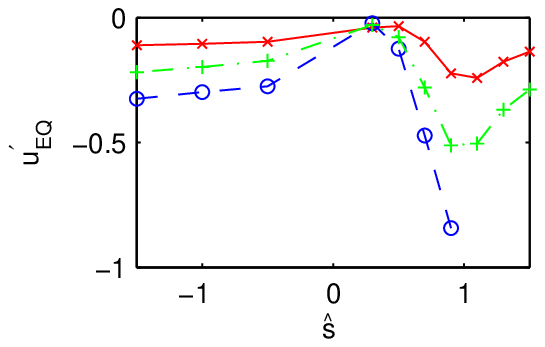}
\includegraphics[width=0.3\linewidth]{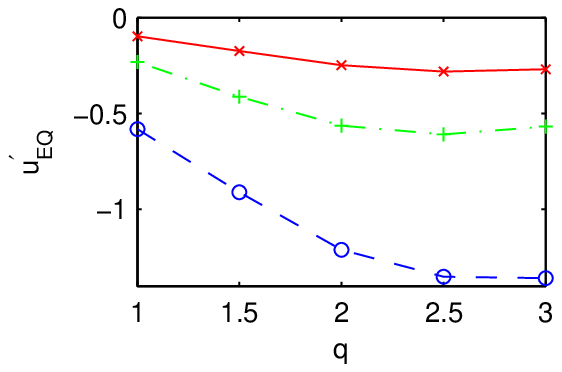}
\includegraphics[width=0.3\linewidth]{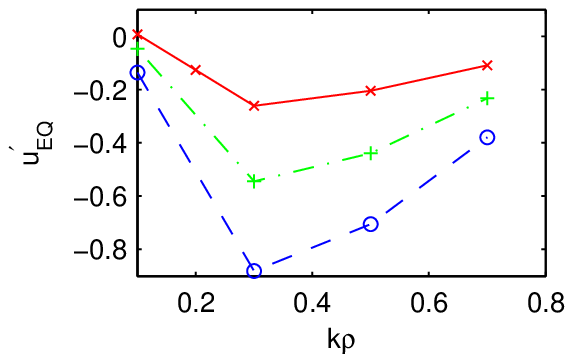}
\includegraphics[width=0.3\linewidth]{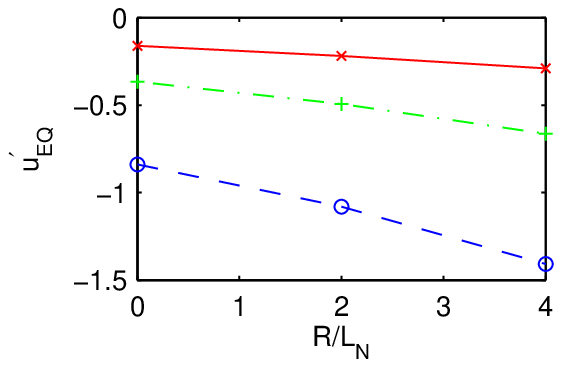}
\includegraphics[width=0.3\linewidth]{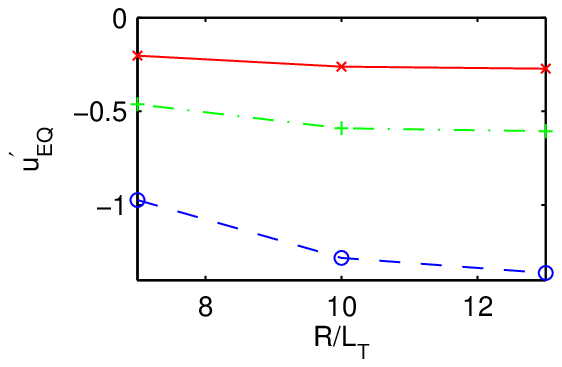}
\caption{
Equilibrium value of $u^\prime$ as a function of different plasma parameters. From left to right: magnetic shear ($\hat s$), 
safety factor ($q$), poloidal wave vector ($k\rho$) density gradient length $R/L_N$ and temperature gradient length $R/L_T$. 
The three curves that are shown in each of the graphs correspond to different values of the inverse aspect ratio: $\epsilon = 0.16$
solid (red) line with symbol 'x', $\epsilon = 0.08$ dash-dotted (green) line with symbol '+', and $\epsilon = 0.05$ dashed (blue) line 
with symbol 'o'. }
\label{parameter}
\end{figure*}

Fig. \ref{parameter} shows the momentum transport for different plasma parameters (with $\rho_* = 2.5\cdot 10^{-3}$). 
For all parameters the momentum transport due to the $\rho_*$ terms leads to a negative $u^\prime_{EQ}$ as predicted by 
Eq.~\eqref{uprim_analytic}. 
Assuming the rotation at the plasma edge is zero, a counter-current rotation is then generated in the core.
It can be seen that $u^\prime_{EQ}$ increases with decreasing $\epsilon$, in agreement with the prediction of the analytic model. 
Also the increase with decreasing $k_\theta \rho$ predicted by this equation is found at 
sufficiently large $k_\theta \rho$. Below $k_\theta \rho = 0.3$ however, the magnitude of the flux decreases with decreasing $k_\theta 
\rho$. Because also $\chi_\varphi$ decreases \cite{PEE06} this represents a strong decrease in the momentum flux due to the 
finite $\rho_*$ terms. This decrease in the flux might be related to a larger extension of the mode along the surface (i.e. a larger 
$L_\theta$ in Eq.~(\ref{uprim_analytic})), but the exact interpretation is at present unknown.
The increase of the momentum flux with safety factor and magnetic shear can be understood 
through a stronger localisation of the mode (smaller $L_\theta$). 
For the lowest $\epsilon$, no data of $u^\prime_{EQ}$ is shown for the highest values of the magnetic shear, because at higher 
shear the dominant instability is the off-axis ITG \cite{MIG13} (this mode also sets in for $k_\theta \rho > 0.7$). 
In the lowest order local limit two modes exist (one shifted towards positive $s$ and one towards negative $s$) that have equal 
growth rates, and drive an equally large momentum flux in the opposite direction resulting in a zero momentum flux on average. 
Even a small symmetry breaking results in one of these modes to be more unstable than the other. 
The time integration projects out only the most unstable mode that than drives a large momentum flux which magnitude no longer 
depends on the magnitude of the symmetry breaking term. 
This case is interesting, but beyond the scope of the present paper.

\section{DISCUSSION}

In this paper the effect of finite $\rho_*$ terms on the toroidal momentum transport is investigated.  
Essentially, higher order $\rho_*$ terms connected with the derivative towards the 'parallel' coordinate $s$ have been investigated.
In the ballooning representation, these are the terms connected with the derivative of the ballooning envelope, rather than the eikonal.
The model equations studied in this paper are homogeneous in the radial direction. 
Finite $\rho_*$ effects due to the radial profiles are, therefore, not included.  

This paper discusses only the quasi-linear theory, and assumes an adiabatic electron response. 
Clearly, further study is required to assess the magnitude of the effect under experimentally relevant conditions. 
Although the momentum is mainly carried by the ions, kinetic electrons as well as finite beta effects have in the past been found 
to have a significant impact on the momentum flux \cite{PEE09d,KLU09,HEI11}. 
From the analytic model one might, for instance, expect that the result is sensitive to the extension of the mode along the field 
line. 
Finally, only non-linear simulations can accurately determine the momentum fluxes. 
These studies are left to future work. 

H-mode plasmas have been reported to rotate mostly in the co-current direction \cite{ric08}, but it appears that both co- and 
counter current rotation can occur if the plasma edge does not rotate strongly\cite{ang11}. 
Experimentally, a transition from co- to counter current rotation is then observed when the density (or density gradient) exceeds a 
threshold value \cite{bor06,ric11,ang11}.    
In our model the momentum flux due to the higher order parallel derivatives is directed such that the plasma core will rotate more strongly in the 
counter-current direction.

\end{document}